\documentclass[pra,twocolumn,showpacs,superscriptaddress]{revtex4}
\usepackage[latin1]{inputenc}
\usepackage{graphicx}
\usepackage{amssymb}
\usepackage{amsmath}
\usepackage{xspace}
\usepackage{bm}
\usepackage{times}
\usepackage{color}

\newfont{\tensy}{cmsy10}

\newcommand{\ie}[0]{i.e.\@\xspace}
\newcommand{\eg}[0]{e.g.\@\xspace}
\newcommand{\etal}[0]{et al.\@\xspace}

\newcommand{\nag}{{\phantom{\dag}}}

\newcommand{\las}[0]{\langle}
\newcommand{\ras}[0]{\rangle}

\renewcommand{\tilde}[1]{\widetilde{#1}}

\begin{document}

\title{Dynamical critical exponent of the Jaynes-Cummings-Hubbard model}

\date{\today}

\author{M. Hohenadler} \affiliation{Institut f\"ur Theoretische Physik und
  Astrophysik, Universit\"at W\"urzburg, 97074 W\"urzburg, Germany}

\author{M. Aichhorn} \affiliation{Institut f\"ur Theoretische Physik --
  Computational Physics, TU Graz, 8010 Graz, Austria}

\author{S. Schmidt} \affiliation{Institut f\"ur Theoretische Physik,
  ETH Zurich, 8093 Z\"urich, Switzerland}

\author{L. Pollet} \affiliation{Institut f\"ur Theoretische Physik,
  ETH Zurich, 8093 Z\"urich, Switzerland}

\begin{abstract}
  An array of high-$Q$ electromagnetic resonators coupled to qubits gives rise
  to the Jaynes-Cummings-Hubbard model describing a superfluid to Mott
  insulator transition of lattice polaritons. From mean-field and strong
  coupling expansions, the critical properties of the model are expected to
  be identical to the scalar Bose-Hubbard model.  A recent Monte Carlo study
  of the superfluid density on the square lattice suggested that this does
  not hold for the fixed-density transition through the Mott lobe
  tip. Instead, mean-field behavior with a dynamical critical exponent $z=2$
  was found. We perform large-scale quantum Monte Carlo simulations to
  investigate the critical behavior of the superfluid density and the
  compressibility. We find $z=1$ at the tip of the insulating lobe. Hence
  the transition falls in the 3D $XY$ universality class, analogous to the
  Bose-Hubbard model.
\end{abstract}


\maketitle

\paragraph*{Introduction}
A remarkable success of theoretical physics is the development of a theory of
phase transitions which provides a unified description of apparently distinct
systems in terms of a few different universality classes characterized by
universal critical exponents.  The universality of a transition is completely
determined by the dimension of the system, the order parameter
and the range of interactions.  An interesting example is the Mott insulator
to superfluid (MI-SF) transition of lattice bosons, first studied in the
framework of the Bose-Hubbard model (BHM) \cite{PhysRevB.40.546}.  When
driven by density fluctuations, the MI-SF transition is known to be in the
universality class of the dilute Bose gas with a dynamical critical exponent
$z=2$ \cite{PhysRevB.40.546,alet:024513}.  However, when the density of bosons is kept
fixed during the transition its universality class changes and corresponds to
the (2+1)D $XY$ model with $z=1$
\cite{PhysRevB.40.546,PhysRevB.59.12184,CaSa.GuSo.Pr.Sv.07}.

The experimental realization of the BHM with ultra-cold atoms in optical
lattices \cite{Gr.Ma.Es.Ha.Bl.02} nearly a decade ago has opened up a fast
growing and versatile field of condensed matter physics.  More recently,
the realization of Bose-Einstein condensation of weakly interacting
polaritons, \ie, quasiparticles that form when light strongly interacts with
matter, in high-$Q$ cavities \cite{KaRiKuBa06} has triggered an immense
interest in realizing condensed matter systems with photonic systems.  A
recent theoretical focus has been on the MI-SF transition of polaritons
\cite{GrTaCoHo06,HaBrPl06,AnSaBo07,Ai.Ho.Ta.Li.08,Ro.Fa.07,Zh.Sa.Ue.08,Ko.LH.09,Sc.Bl.09,PhysRevLett.104.216402,PhysRevB.82.045126}.
The Jaynes-Cummings-Hubbard model (JCHM) has been introduced to describe
such a quantum phase transition of light in an array of coupled quantum
electrodynamics (QED) cavities,
each containing a single photonic mode interacting with a two-level system
(qubit) \cite{GrTaCoHo06,HaBrPl06,AnSaBo07}.  The competition between strong
photon-qubit coupling, giving rise to an effective photonic repulsion
(localization), and the photon hopping between cavities (delocalization)
leads to a phase diagram featuring Mott lobes reminiscent of those of
ultracold atoms in optical lattices as described by the BHM.  The realization
of the JCHM has been proposed in various solid-state or quantum-optical systems,
\eg, with nitrogen-vacancy (NV) centers in diamond \cite{GrTaCoHo06,HaBrPl06,AnSaBo07}, quantum
dot excitons \cite{Na.Ut.Ti.Ya.07}, superconducting qubits \cite{Ko.LH.09}
and trapped ions \cite{Fleischhauer-ions}.  Device integration, high
tunability and individual addressability of each cavity make wide parameter
regimes easily accessible. Cavity or circuit QED arrays thus constitute one of
the most promising architectures for quantum information processing and offer
the possibility to study fundamental questions of interacting quantum
systems.  For a recent review of many-body physics in coupled-cavity arrays
see Ref.~\cite{Ha.Br.Pl.08}.

The phase diagram and excitation spectra of the JCHM have been accurately
determined by analytical
\cite{GrTaCoHo06,Sc.Bl.09,Ko.LH.09,PhysRevLett.104.216402} as well as
numerical methods
\cite{HaBrPl06,AnSaBo07,Ai.Ho.Ta.Li.08,Ro.Fa.07,Zh.Sa.Ue.08,Pi.Ev.Ho.09,PhysRevB.82.045126}.
However, an intriguing open question concerns the universality class of the
fixed-density transition of the JCHM.  Quantum Monte Carlo (QMC) calculations
of the superfluid density \cite{Zh.Sa.Ue.08} suggest that the universality
class at the tip of the Mott lobe of the JCHM is different (namely mean-field
like) from the BHM.  On the contrary, strong-coupling expansion
\cite{Sc.Bl.09,PhysRevLett.104.216402} and an effective action approach \cite{Ko.LH.09}
show that the same critical theory applies to both models.  The discrepancy
with the QMC predictions is very surprising, especially since the analytical
arguments for arriving at a critical theory are similar to the BHM and are well
understood. One would therefore expect the same scaling behavior, \eg, of the
superfluid density, for the BHM and the JCHM.
 
In this Rapid Communication we resolve the controversy between analytical and numerical
findings and present results from extensive QMC simulations on the
two-dimensional (2D) square lattice. We go beyond previous studies in two
significant ways: (i) by using much larger system sizes, and (ii) by studying
both the superfluid density and the compressibility.

\begin{figure}[t]
  \centering
  \includegraphics[width=0.45\textwidth]{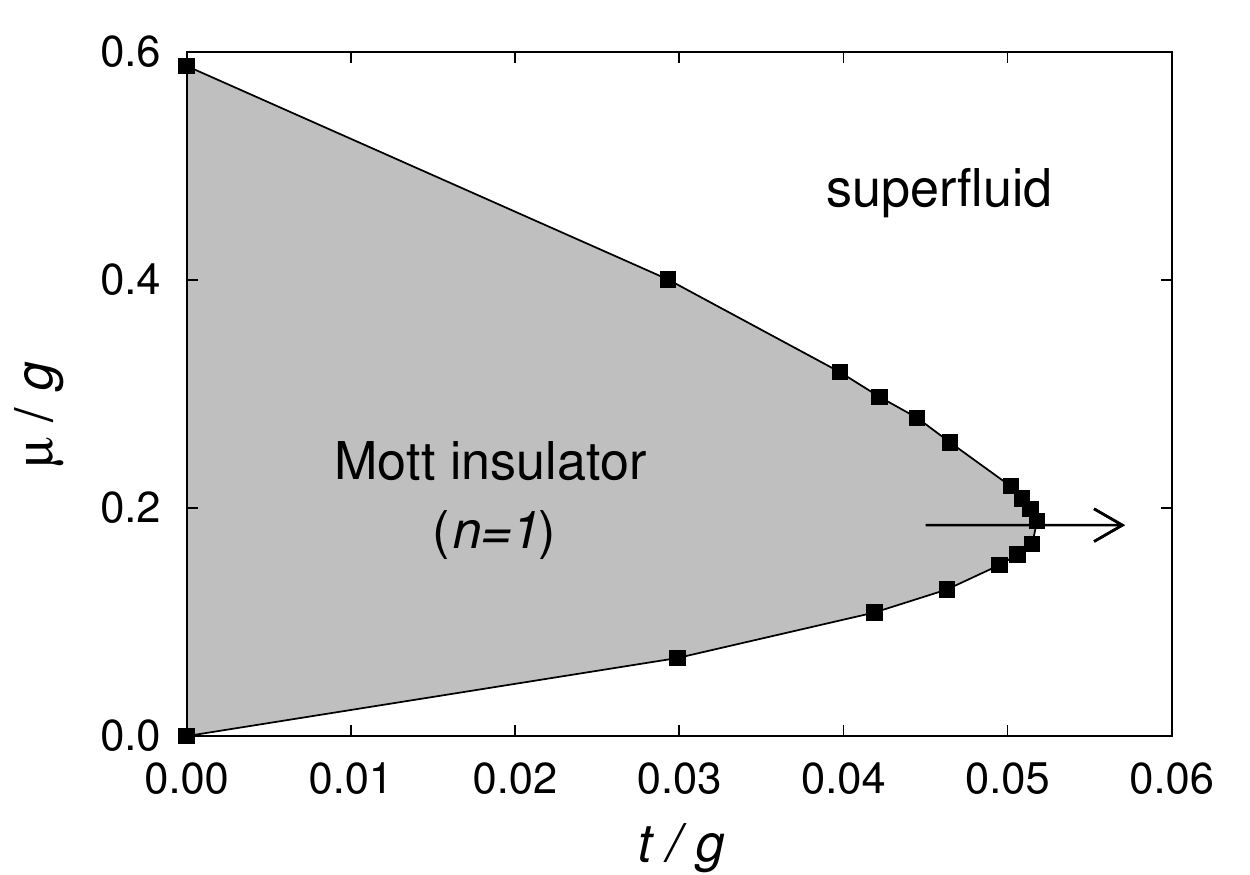}
  \caption{\label{fig:2d_qmc}%
    Phase diagram of the JCHM with zero detuning on the 2D square lattice as
    obtained from QMC simulations \cite{Zh.Sa.Ue.08}. Shown is the first Mott
    lobe with filling factor $n=1$. The arrow indicates the fixed-density
    transition through the tip of the lobe.}
\end{figure}

\paragraph*{Model}
The JCHM is defined by the Hamiltonian
\begin{align}\label{eq:ham}
  H= \sum_i h^\text{\rm JC}_i -t \sum_{\langle ij\rangle } (a^\dag_i
  a^\nag_j+\text{H.c.}) - \mu \hat{N}\,,
\end{align}
with the JC Hamiltonian for site $i$ \cite{Ja.CU.63}
\begin{align}\label{eq:ham3}
  h^\text{JC}_i = \omega_\text{p} a^\dag_i a^\nag_i + \omega_\text{q}
  \sigma^+_i \sigma^-_i + g \left(\sigma^+_i a^\nag_i + \sigma^-_i
    a^\dag_i\right) \,,\nonumber
\end{align}
and the total number of polaritons $\hat{N}=\sum_i (a^\dag_i a^\nag_i +
\sigma^+_i\sigma^-_i)$ (combined number of photons and qubit excitations),
which is a conserved quantity and can be controlled via the chemical
potential $\mu$.  Here, $\omega_\text{p}$ ($\omega_\text{q}$) is the energy
of photons (qubits).  The spin operators $\sigma^\pm$ describe intrasite
transitions between the two qubit levels induced by emission or absorption of
a photon with rate $g$ (light-matter coupling). The second term in
(\ref{eq:ham}) describes photon transfer between nearest neighbor sites $i$
and $j$ with hopping amplitude $t$.  The polariton density (filling factor) is
denoted as $n\equiv\las\hat{n}\ras=\las\hat{N}\ras/L^2$, where $L^2$ is the
number of lattice sites. We use $g$ as the unit of energy, and consider
$\omega_\text{p}=\omega_\text{q}=1$ (resonance condition).

Similar to the BHM, the ground-state phase diagram of the JCHM consists of a
series of Mott-insulating lobes with integer filling factor $n$. The extent
of the lowest Mott lobe with $n=1$ for the 2D square lattice considered here
is known from QMC simulations \cite{Zh.Sa.Ue.08}, see Fig.~\ref{fig:2d_qmc}.
The generic MI-SF transition in the 2D BHM caused by
density fluctuations is known to be of the mean-field type with a dynamical
critical exponent $z=2$ \cite{PhysRevB.40.546,alet:024513}.  However,
when increasing the hopping $t$ along a path that leads through the tip of
the Mott lobe while keeping the polariton density constant, the transition is
driven by phase fluctuations. This special case belongs to the universality
class of the (2+1)D $XY$ model with $z=1$
\cite{PhysRevB.40.546,CaSa.GuSo.Pr.Sv.07}.  This means that particle and hole
excitations become symmetric near the tip of the lobe and that the critical
field theory describing this transition is relativistic with equal scaling of
space and time directions. The location of the lobe tip  has been determined
as $\mu/g=0.185(5)$, $t_\text{c}/g=0.05201(10)$ \cite{Zh.Sa.Ue.08}. 
For the 2D JCHM, previous QMC simulations have
confirmed $z=2$ for the generic, density-driven transition. However, the same
QMC simulations also suggest that the special $z=1$ point at the tip of the
lobe in the BHM is absent in the JCHM. On the other hand, analytical
calculations of the excitation spectra \cite{Sc.Bl.09,PhysRevLett.104.216402}
and general symmetry arguments based on the gauge invariance of the theory
\cite{Ko.LH.09} predict a special point with $z=1$ at the lobe tip in the
JCHM as well.

\paragraph*{Method}
In order to resolve this controversy and to determine the dynamical critical
exponent $z$, we compute the finite-size scaling behavior of the superfluid
density $\rho_\text{s}$ and the compressibility $\kappa$. Scaling relations
for both quantities are known from Ref.~\cite{PhysRevB.40.546}. The superfluid
density is related to fluctuations of the winding number $W$ in QMC
simulations via \cite{PhysRevB.36.8343}
\begin{equation}
  \rho_\text{s} = \frac{\las W^2\ras}{\beta D L^{D-2}}\,.
  \label{eq:rhosf_def}
\end{equation}
The finite-size scaling form of $\rho_\text{s}$ reads
\begin{equation}
  \rho_\text{s} =
  L^{2-D-z}\tilde{\rho}_\text{s}[(t-t_\text{c})L^{1/\nu},\beta/L^z]\,,
  \label{eq:rhosf_FSS}
\end{equation}
where $\nu$ denotes the correlation length exponent.
Fixing the ratio $\alpha=\beta/L^z$, the quantity
\begin{equation}
  L^{D-2+z} \rho_\text{s} [(t-t_\text{c})L^{1/\nu},\alpha]
\end{equation}
depends only on the distance from the critical point,
$(t-t_\text{c})L^{1/\nu}$, so that at $t=t_\text{c}$ curves for different $L$
as a function of $t$ intersect in a single point. This allows to determine
the critical value $t_\text{c}$ for the MI-SF transition. Plotting $L^{D+z-2}
\rho_\text{s}$ as a function of $(t-t_\text{c})L^{1/\nu}$ should lead to a
scaling collapse of results for different $L$.

A second observable of interest is the compressibility $\kappa = \partial
n/\partial \mu$, which, by the fluctuation-dissipation theorem, can also be
expressed in terms of number fluctuations, \ie,
\begin{equation}
  \kappa = \beta \left(\las \hat{n}^2 \ras - \las \hat{n} \ras^2 \right)
  \,,
\end{equation}
with the scaling form
\begin{equation}\label{eq:kappa}
  \kappa =
  L^{z-D}\tilde{\kappa}[(t-t_\text{c})L^{1/\nu},\alpha]
\end{equation}
implying that $L^{D-z}\kappa$ should be independent of $L$ at $t_\text{c}$.

For the calculation of $\rho_\text{s}$ and $\kappa$, the
Hamiltonian~(\ref{eq:ham}) is simulated using world lines in the
stochastic series expansion (SSE) representation
\cite{Pi.Ev.Ho.09,ALPS_I}. This is the same method as previously used by Zhao
\etal \cite{Zh.Sa.Ue.08}.  We have used the ALPS 1.3 implementation
\cite{ALPS_I} of the SSE with directed loop updates
\cite{SySa02,ALPS_DIRLOOP}.  In the vicinity of the MI-SF boundary of the
first Mott lobe, allowing a maximum of six photons respectively polaritons per
site is sufficient to eliminate any noticeable error. We considered $L\times
L$ square lattices. The inverse temperature $\beta$ was chosen as $\beta
g=2L$ for $z=1$ and $\beta g=L^2/4$ for $z=2$.  We also performed QMC
simulations using the worm algorithm \cite{worm} in the path integral
representation following the implementation of Ref.~\cite{worm_lode}.  No cutoff
in the polariton number is needed in this case. The results on up to
$64\times64$ lattices (not shown) are in full agreement with our SSE data.

\begin{figure}[t]
  \centering
  \includegraphics[width=0.45\textwidth]{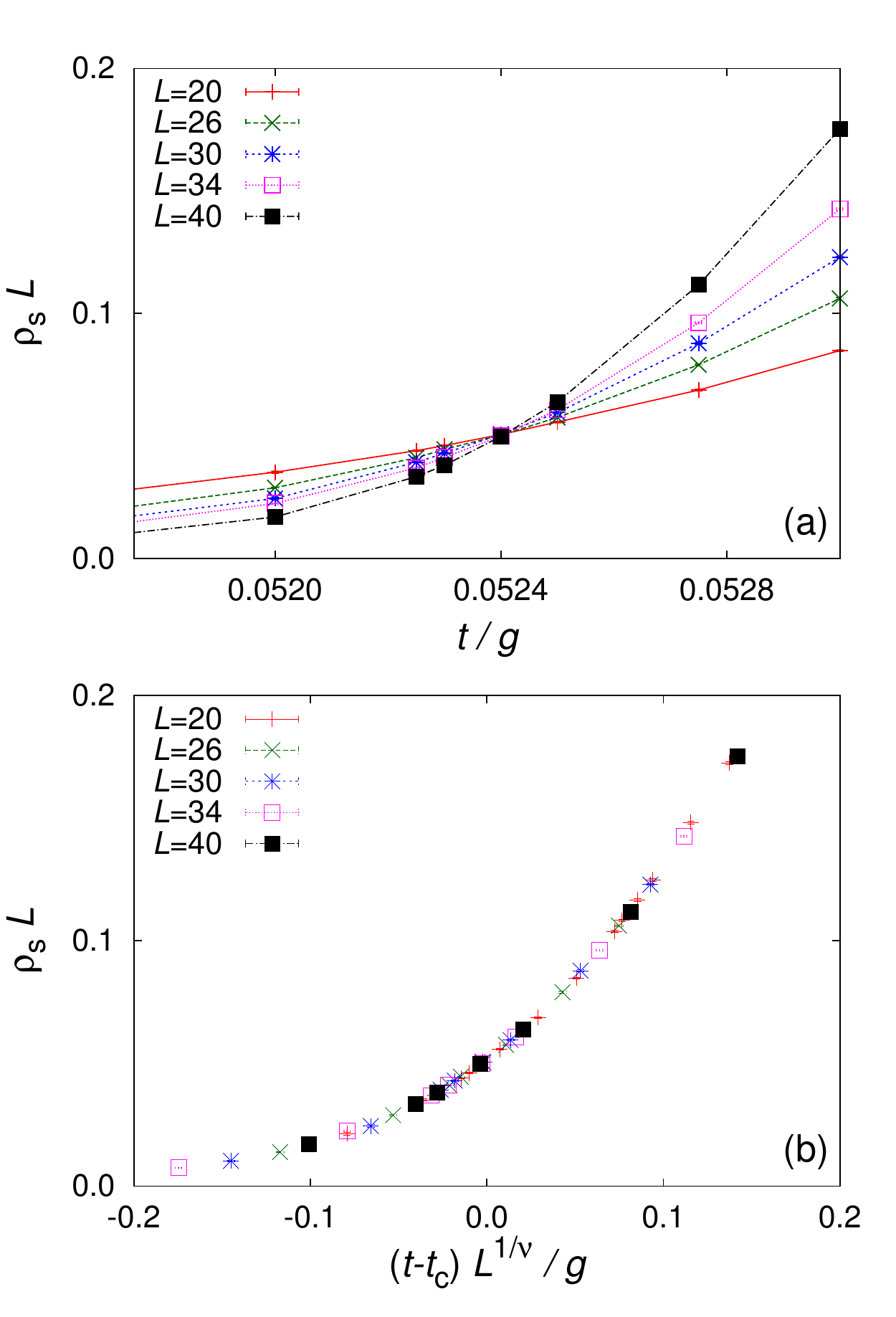}
  \caption{\label{fig:2d_qmc_scaling_z1rho}%
    (Color online) Scaling of the superfluid density $\rho_\text{s}$ across
    the fixed-density transition at $\mu/g=0.185$ \cite{Zh.Sa.Ue.08} using
    $L\times L$ square lattices and $\beta g=2L$. The intersect of $L
    \rho_\text{s}$ for different lattice sizes $L$ in a single point in panel
    (a) is evidence for a dynamical critical exponent $z=1$, and defines the
    critical point at $t_\text{c}/g=0.05242(1)$. (b) Scaling collapse using
    $t_\text{c}/g=0.05242$ and $\nu=0.6715$ \cite{CaSa.GuSo.Pr.Sv.07}.}
\end{figure}

\begin{figure}[t]
  \centering
  \includegraphics[width=0.45\textwidth]{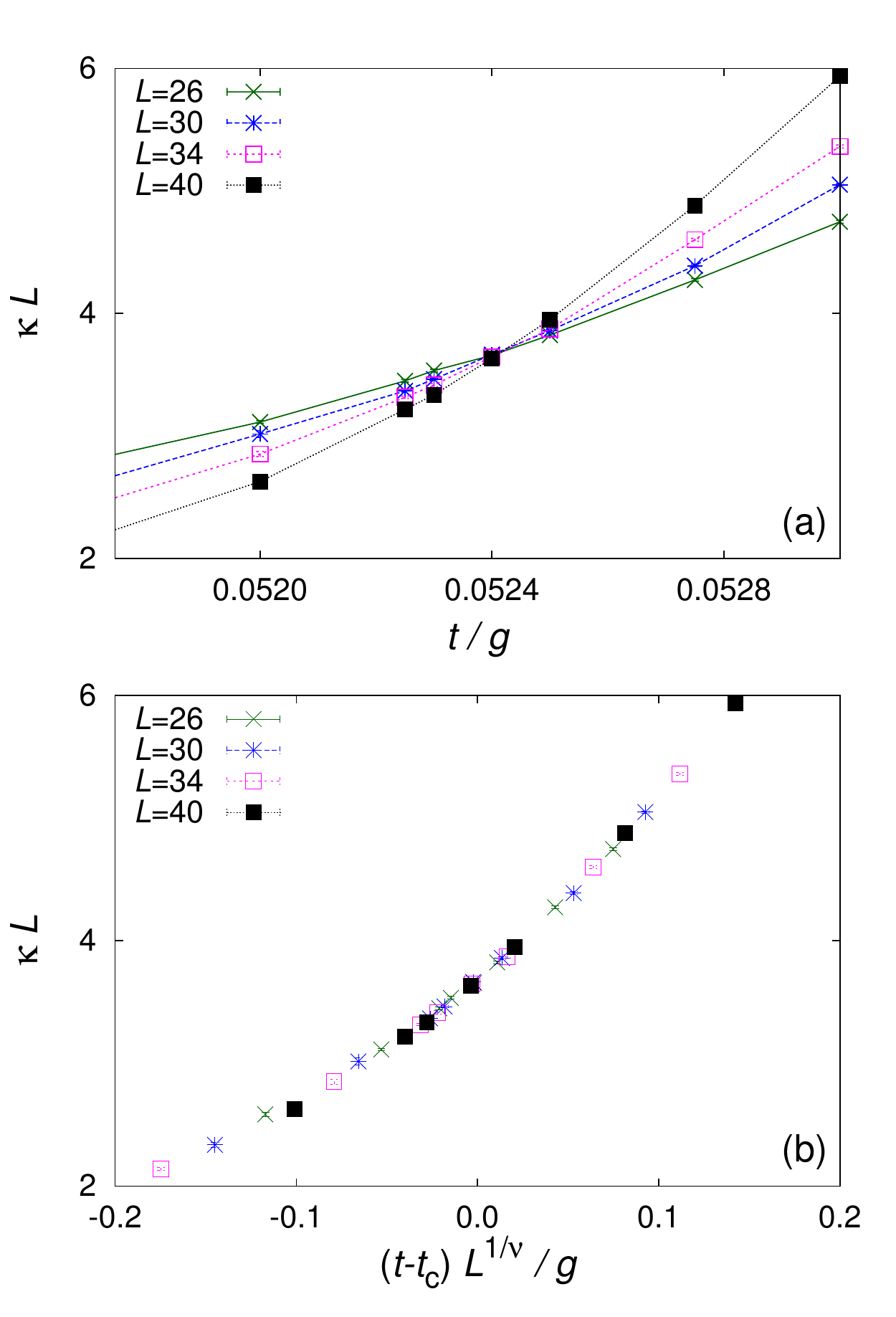}
  \caption{\label{fig:2d_qmc_scaling_z1kappa}%
    (Color online) Scaling of the compressibility $\kappa$ for the same
    parameters as in Fig.~\ref{fig:2d_qmc_scaling_z1rho}.  The intersect in
    (a) is consistent with $z=1$ and occurs at the same value of the critical
    hopping strength $t_\text{c}/g=0.05242(1)$ as in
    Fig.~\ref{fig:2d_qmc_scaling_z1rho}. (b) Scaling collapse using
    $t_\text{c}/g=0.05242$ and $\nu=0.6715$ \cite{CaSa.GuSo.Pr.Sv.07}.}
\end{figure}

The conclusion of $z=2$ scaling for the whole MI-SF phase boundary in previous
work was based on results for the superfluid density on lattice sizes up to
$22\times22$ \cite{Zh.Sa.Ue.08}. Here we present data for both the superfluid
density and the compressibility, using much larger system sizes up to
$40\times40$. The existence of a $z=1$ critical point should also be visible
in numerical simulations in a finite range of parameters around the lobe tip
\cite{Zh.Sa.Ue.08}. Hence, a grand-canonical algorithm with a suitably
chosen chemical potential (here $\mu/g=0.185$ \cite{Zh.Sa.Ue.08}) can be used,
as confirmed by the results below.

\paragraph*{Results}
First, we consider the scaling behavior of $\rho_\text{s}$ and $\kappa$
assuming $z=1$.  In that case fluctuations in space and (imaginary) time are
isotropic and the winding number in space (leading to a nonzero superfluid
stiffness) scales the same way as the winding number in imaginary time
(leading to a nonzero compressibility).  Results for $z=1$ with the aspect
ratio $\beta g/L=\text{const.}=2$ are shown in
Figs.~\ref{fig:2d_qmc_scaling_z1rho} and~\ref{fig:2d_qmc_scaling_z1kappa}.

Figure~\ref{fig:2d_qmc_scaling_z1rho}(a) shows the rescaled superfluid
density $\rho_\text{s}L$ as a function of the hopping strength $t/g$ for
system sizes ranging from $20\times20$ to $40\times40$. The intersect of the
curves leads to the estimate of the critical hopping strength
$t_\text{c}/g=0.05241(1)$. The previous estimate was
$t_\text{c}/g=0.05201(10)$ \cite{Zh.Sa.Ue.08}. Our value of $t_\text{c}$,
together with the correlation length exponent $\nu=0.6715$ (as found
numerically for the BHM \cite{CaSa.GuSo.Pr.Sv.07}), leads to a clean scaling
collapse in Fig.~\ref{fig:2d_qmc_scaling_z1rho}(b). We also observe that we
need bigger system sizes for the JCHM as compared to the BHM in order to see
a clear scaling collapse.

Figure~\ref{fig:2d_qmc_scaling_z1kappa} shows a similar analysis for the
compressibility $\kappa$. The value of $t_\text{c}$ from the intersect in
Fig.~\ref{fig:2d_qmc_scaling_z1kappa}(a) coincides within errorbars with the
value obtained from the superfluid density. Again, we find a very clean
scaling collapse in Fig.~\ref{fig:2d_qmc_scaling_z1kappa}(b).  Our results
are thus fully consistent with $z=1$.

\begin{figure}
  \centering
  \includegraphics[width=0.45\textwidth]{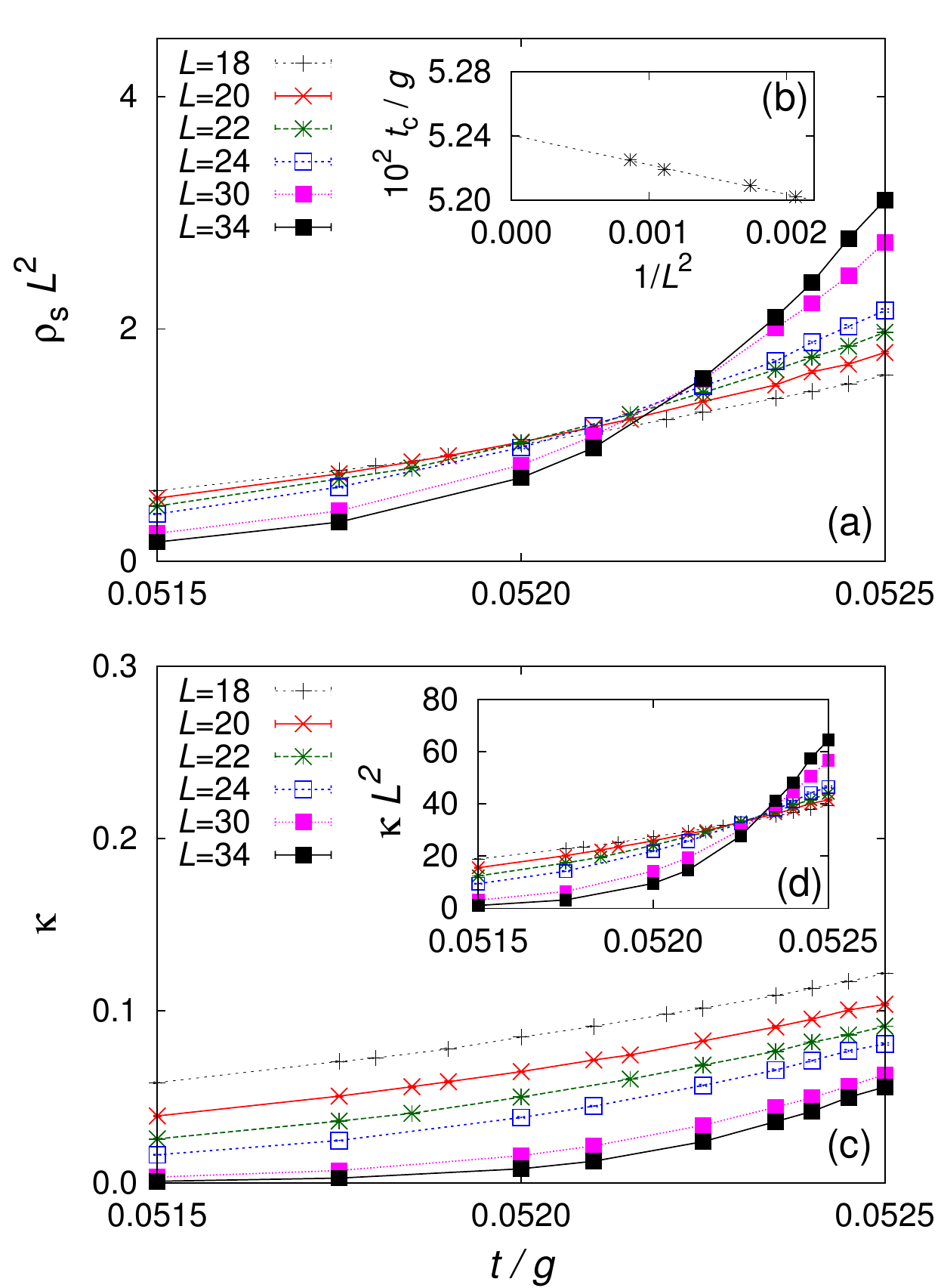}
  \caption{\label{fig:2d_qmc_scaling_z2}%
    (Color online) Scaling of the superfluid density $\rho_\text{s}$ and
    compressibility $\kappa$ across the fixed-density transition at
    $\mu/g=0.185$ \cite{Zh.Sa.Ue.08} using $\beta g=L^2/4$.  In (a), the
    intersect of curves for consecutive system sizes [shown in panel (b)]
    approaches the $t_\text{c}$ obtained in
    Figs.~\ref{fig:2d_qmc_scaling_z1rho}
    and~\ref{fig:2d_qmc_scaling_z1kappa}. Panel (c) shows the compressibility
    for the same parameters, with no intersection in the vicinity of
    $t_\text{c}$. Panel (d) illustrates the intersect obtained when plotting
    $\kappa L^2$.}
\end{figure}

In Ref.~\cite{Zh.Sa.Ue.08}, an apparent scaling collapse for the superfluid
density was found assuming $z=2$.  In Fig.~\ref{fig:2d_qmc_scaling_z2} we
therefore present our results for the case $z=2$ with $\beta
g/L^2=\text{const.}=1/4$.  We point out that we do not find any contradiction
between our numerical data and those of Ref.~\cite{Zh.Sa.Ue.08} when considering
the same system sizes. The curves for the superfluid density, shown in the
Fig.~\ref{fig:2d_qmc_scaling_z2}(a), seem to cross in a single point and one
may be tempted to think that $z=2$ applies equally well \cite{Zh.Sa.Ue.08}.
However, differences clearly show up for larger system sizes.
Figure~\ref{fig:2d_qmc_scaling_z2}(b) shows finite-size scaling corrections
to the critical value of the hopping strength that go as $1/L^2$. The
extrapolation of those intersection points yields a critical value for the
hopping strength that is within error bars the same as the one found for
$z=1$ scaling. The estimate of $t_\text{c}$ for smaller $L$ matches the
result of Ref.~\cite{Zh.Sa.Ue.08}.  These observations can be understood from
Eq.~(\ref{eq:rhosf_def}): The winding number, which is an integer, is in 2D
given by $\langle W^2\rangle \sim \rho_s \beta$, and hence the leading term in
the finite size scaling for the superfluid density cannot distinguish between
$z=1$ and $z=2$.  The small subleading corrections we see in
Fig.~\ref{fig:2d_qmc_scaling_z2}(b) (while we were unable to resolve any such
drifts in the $z=1$ scenario within the resolution of our numerical data)
therefore provide  further evidence that $z=1$ is correct. Using the full
range of system sizes available here, no scaling collapse is achieved
assuming $z=2$.

An even stronger distinction between $z=1$ and $z=2$ can be made using
results for the compressibility. To this end, we plot in
Fig.~\ref{fig:2d_qmc_scaling_z2}(b) $\kappa L^0=\kappa$ as appropriate for
$z=2$---cf. Eq.~(\ref{eq:kappa}). The absence of a crossing point on the scale
of the plot suggests that there are strong corrections to the scaling
assumption, making $z=2$ very unlikely. In contrast, $\kappa$ scales as
expected for the generic transition of the JCHM (not shown).

Finally, if $z=1$ is correct, relativistic invariance allows us to interpret
the imaginary time axis as a spatial axis, and vice versa. Consequently,
$\kappa L^2$ (with $\beta \sim L^2$ scaling) should behave identically to
$\rho L^2$. This is shown in Fig.~\ref{fig:2d_qmc_scaling_z2}(d).  We
therefore conclude that $z = 1$ must be true, and we see no physical reason
to investigate other scenarios such as theories with a fractional $z$.

\paragraph*{Summary}
We have shown on the basis of finite-size scaling of the superfluid density
and the compressibility that the fixed-density Mott-insulator-superfluid transition in
the 2D Jaynes-Cummings-Hubbard model falls into the 3D $XY$ universality
class. The critical behavior is thus identical to the corresponding
transition in the Bose-Hubbard model.

\paragraph*{Acknowledgments}
We are grateful to G.~Blatter, J.~Keeling, K.~Le Hur, P.~Pippan,
N.~V. Prokof'ev, and M.~Troyer for valuable discussions.  MH was supported by
the DFG FG1162, and acknowledges the hospitality of ETH Zurich. LP was
supported by the Swiss National Science Foundation under Grant
No. PZ00P2-131892/1.  Part of the simulations were performed on the Brutus
cluster at ETH Zurich.


\begin{thebibliography}{10}

\bibitem{PhysRevB.40.546}
M.~P.~A. Fisher, P.~B. Weichman, G. Grinstein, and D.~S. Fisher, Phys. Rev. B
  {\bf 40},  546  (1989).

\bibitem{alet:024513}
F. Alet and E.~S. Sorensen, Phys. Rev. B {\bf 70},  024513  (2004).

\bibitem{PhysRevB.59.12184}
N. Elstner and H. Monien, Phys. Rev. B {\bf 59},  12184  (1999).

\bibitem{CaSa.GuSo.Pr.Sv.07}
B. {Capogrosso-Sansone}, S.~G. S\"oyler, N. Prokof'ev, and B. Svistunov, Phys.
  Rev. A {\bf 77},  015602  (2008).

\bibitem{Gr.Ma.Es.Ha.Bl.02}
M. Greiner {\it et~al.}, Nature (London) {\bf 415},  39  (2002).

\bibitem{KaRiKuBa06}
J. Kasprzak {\it et~al.}, Nature (London) {\bf 443},  409  (2006).

\bibitem{GrTaCoHo06}
A.~D. Greentree, C. Tahan, J.~H. Cole, and L.~C.~L. Hollenberg, Nat. Phys. {\bf
  2},  856  (2006).

\bibitem{HaBrPl06}
M.~J. Hartmann, F.~G. S.~L. Brand{\~{a}}o, and M.~B. Plenio, Nat. Phys. {\bf
  2},  849  (2006).

\bibitem{AnSaBo07}
D.~G. Angelakis, M.~F. Santos, and S. Bose, Phys. Rev. A {\bf 76},  031805
  (2007).

\bibitem{Ai.Ho.Ta.Li.08}
M. Aichhorn, M. Hohenadler, C. Tahan, and P.~B. Littlewood, Phys. Rev. Lett.
  {\bf 100},  216401  (2008).

\bibitem{Ro.Fa.07}
D. Rossini and R. Fazio, Phys. Rev. Lett. {\bf 99},  186401  (2007).

\bibitem{Zh.Sa.Ue.08}
J. Zhao, A.~W. Sandvik, and K. Ueda, arXiv:0806.3603  (2008).

\bibitem{Ko.LH.09}
J. Koch and K. Le~Hur, Phys. Rev. A {\bf 80},  023811  (2009).

\bibitem{Sc.Bl.09}
S. Schmidt and G. Blatter, Phys. Rev. Lett. {\bf 103},  086403  (2009).

\bibitem{PhysRevLett.104.216402}
S. Schmidt and G. Blatter, Phys. Rev. Lett. {\bf 104},  216402  (2010).

\bibitem{PhysRevB.82.045126}
M. Knap, E. Arrigoni, and W. von~der Linden, Phys. Rev. B {\bf 82},  045126
  (2010).

\bibitem{Na.Ut.Ti.Ya.07}
N. Na, S. Utsunomiya, L. Tian, and Y. Yamamoto, Phys. Rev. A {\bf 77},  031803
  (2008).

\bibitem{Fleischhauer-ions}
P.~A. Ivanov {\it et~al.}, Phys. Rev. A {\bf 80},  060301  (2009).

\bibitem{Ha.Br.Pl.08}
M. Hartmann, F. Brandao, and M. Plenio, Laser \& Photonics Review {\bf 2},  527
   (2008).

\bibitem{Pi.Ev.Ho.09}
P. Pippan, H.~G. Evertz, and M. Hohenadler, Phys. Rev. A {\bf 80},  033612
  (2009).

\bibitem{Ja.CU.63}
E.~T. Jaynes and F.~W. Cummings, Proc. IEEE {\bf 51},  89  (1963).

\bibitem{PhysRevB.36.8343}
E.~L. Pollock and D.~M. Ceperley, Phys. Rev. B {\bf 36},  8343  (1987).

\bibitem{ALPS_I}
A. Albuquerque {\it et~al.}, J. Magn. Magn. Mater. {\bf 310},  1187  (2007).

\bibitem{SySa02}
O.~F. Sylju{\aa}sen and A.~W. Sandvik, Phys. Rev. E {\bf 66},  046701  (2002).

\bibitem{ALPS_DIRLOOP}
F. Alet, S. Wessel, and M. Troyer, Phys. Rev. E {\bf 71},  036706  (2005).

\bibitem{worm}
N.~V. Prokof'ev, B.~V. Svistunov, and I.~S. Tupitsyn, J. Exp. Theor. Phys. {\bf
  87},  310  (1998).

\bibitem{worm_lode}
L. Pollet, K. Van~Houcke, and S.~M.~A. Rombouts, J. Comput. Phys. {\bf 225},
  2249  (2007).

\end{thebibliography}
\end{document}